**This is a submission version.**

**Please check for the latest published version, if any.**

# An Immersive Virtual Reality Serious Game to Enhance Earthquake Behavioral Responses and Post-earthquake Evacuation Preparedness in Buildings

Zhenan Feng, Vicente A. González, Robert Amor, Michael Spearpoint, Jared Thomas, Rafael Sacks, Ruggiero Lovreglio, Guillermo Cabrera-Guerrero

## Abstract

Enhancing the earthquake behavioral responses and post-earthquake evacuation preparedness of building occupants is beneficial to increasing their chances of survival and reducing casualties after the main shock of an earthquake. Traditionally, training approaches such as seminars, posters, videos or drills are applied to enhance preparedness. However, they are not highly engaging and have limited sensory capabilities to mimic life-threatening scenarios for the purpose of training potential participants. Immersive Virtual Reality (IVR) and Serious Games (SG) as innovative digital technologies can be used to create training tools to overcome these limitations. In this study, we propose an IVR SG-based training system to improve earthquake behavioral responses and post-earthquake evacuation preparedness. Auckland City Hospital was chosen as a case study to test our IVR SG training system. A set of learning outcomes based on best evacuation practice has been identified and embedded into several training scenarios of the IVR SG. Hospital staff (healthcare and administrative professionals) and visitors were recruited as participants to be exposed to these training scenarios. Participants' preparedness has been measured along two dimensions: 1) Knowledge about best evacuation practice; 2) Self-efficacy in dealing with earthquake emergencies. Assessment results showed that there was a significant knowledge and self-efficacy increase after the training. And participants acknowledged that it was easy and engaging to learn best evacuation practice knowledge through the IVR SG training system.

**Keywords:** Immersive Virtual Reality; Serious Games; Earthquake Emergencies; Evacuation Training

# 1. Introduction

Earthquakes are commonly experienced disasters across the world. Every year it is estimated that 100 significant earthquakes hit different areas of the world with varying levels of structural and non-structural damage (Coburn, Spence, & Pomonis, 1992; United States Geological Survey, 2018). The structural integrity of buildings can be increased to prevent structural collapse (Ye, Qu, Lu, & Feng, 2008). Besides, proper and immediate behavioral responses during earthquakes and post-earthquake evacuation are key factors in reducing the impacts of non-structural damage (Alexander, 2012; Bernardini, D'Orazio, & Quagliarini, 2016). "Drop, cover and hold" and a list of follow-on behaviors are recommended as best practice in earthquake-prone countries (Mahdavifar, Izadkhah, & Heshmati, 2009; New Zealand Ministry of Civil Defence & Emergency Management, 2015). Different educational approaches have been adopted to foster the recommended behaviors focused on building occupants such as seminars, posters, or videos. However, these educational approaches often have low emotional engagement and lack realistic hazardous situations and so may not lead to a behavioral shift towards best practice (Chittaro & Ranon, 2009). Apart from these educational approaches, building occupants also can receive practical training through evacuation drills. Bernardini et al. (2016) argued that building occupants might have different behavioral responses in evacuation drills in comparison to a real earthquake evacuation process. One possible reason is that evacuation drills are not able to realistically represent actual hazards; thus, this may lead to a reduced impact on learning outcomes and behavioral changes (Lovreglio, Gonzalez, Amor, Spearpoint, Thomas, Trotter et al., 2017). Besides, building occupants often receive no individual feedback indicating how well they conform to best practice after performing evacuation drills. Without feedback for assessment, building occupants may bring inappropriate behaviors into actual earthquake emergencies. As a result, the effectiveness of these teaching and training approaches is limited in terms of their pedagogical outcomes (Lovreglio et al., 2017).

In order to overcome the limitations mentioned above, innovative digital technologies such as Immersive Virtual Reality (IVR) and Serious Games (SGs) have been introduced for teaching and training purposes in recent years (Freina & Ott, 2015). IVR is a technology that can immerse participants in computer-generated virtual environments (LaValle, 2017). By using IVR, more realistic hazards and threats can be simulated and presented to participants in order to provide life-threatening scenarios to be used in training environments. SGs are a form of video games with a pedagogical goal as one of their primary purposes (Wouters, Van der Spek, & Van Oostendorp, 2009). Participants can interact with gaming objects and receive feedback accordingly, increasing engagement and motivation (Wouters, Van Nimwegen, Van Oostendorp, & Van Der Spek, 2013). SGs can assist in the effective development of IVR educational applications. IVR SGs have been widely adopted for professional training such as healthcare (Ma, Jain, & Anderson, 2014), driving (Ihemedu-Steinke, Erbach, Halady, Meixner, & Weber, 2017), and workplace health and safety (Grabowski & Jankowski, 2015). However, applications of IVR SGs for evacuation training are still rare, especially in the domain of earthquake emergencies, as indicated by a recent systematic literature review (Feng, González, Amor, Lovreglio, & Cabrera-Guerrero, 2018).

The objective of this study is to investigate the effectiveness and applicability of IVR SGs as training tools to enhance earthquake immediate behavioral responses and post-earthquake evacuation preparedness. In this study, we propose an IVR SG framework for a hypothetical earthquake emergency occurring at Auckland City Hospital. This assisted in training participants about best evacuation practice according to the New Zealand Civil Defence guidelines (New Zealand Ministry of Civil Defence & Emergency Management, 2015) and Auckland District Health Board Evacuation Plans (Auckland District Health Board, 2009). Individuals' preparedness has been measured by two dimensions: 1) Knowledge about best evacuation practice; 2) Self-efficacy in dealing with earthquake emergencies. A knowledge test was conducted before and immediately after the training in order to assess the effectiveness of immediate knowledge acquisition. A questionnaire was answered before and immediately after the training in order to measure the effectiveness of self-efficacy improvement. Also, a questionnaire including training efficacy and engagement measurements was answered after the training in order to measure the applicability of the IVR SG training system.

This paper provides the background of IVR SGs for emergency preparedness, and national and hospital earthquake response procedures in New Zealand. It then introduces the proposed IVR SG in detail, presents the research methods applied, and reports and discusses the results.

## 2. Background

### 2.1 IVR SGs for emergency preparedness

The use of Serious Games (SGs) for education and training can be traced back to the late twentieth century (Rice, 2007). SGs are identified as video games with serious purposes, such as training, simulation and healthcare, instead of pure entertainment (Michael & Chen, 2005; Susi, Johannesson, & Backlund, 2007). In terms of emergency preparedness, SGs have been used in two different domains: emergency training and behavioral analysis (Almeida, Rossetti, Jacob, Faria, & Coelho, 2017; Capuano & King, 2015). Examples of emergency training include fire evacuation (Rüppel & Schatz, 2011), aviation evacuation (Chittaro, 2012), and earthquake preparedness (Barreto, Prada, Santos, Ferreira, O'Neill, & Oliveira, 2014). These applications target the general public, aiming at increasing their safety knowledge. One key reason to use SGs for training is that participants generally feel more engaged and motivated with training processes as compared to other approaches such as watching videos or attending seminars (Papastergiou, 2009). By involving gaming mechanisms in training participants are able to interact with objects and environments that assist them to focus on learning content and feedback to enhance learning outcomes (Bellotti, Kapralos, Lee, Moreno-Ger, & Berta, 2013). SGs have been suggested as an effective approach to reinforce traditional training approaches (Gao, González, & Yiu, 2019). Apart from their pedagogical use, SGs have been applied to investigate human behaviors. Participants' in-game behaviors can be monitored and recorded by built-in game mechanisms and scripts. By doing so, SGs have the capability to capture each individual's behaviors; and thus, to be applied for behavioral analysis (Chittaro & Ranon, 2009). Kinateder et al. (2014a; 2014b) recorded participants' in-game evacuation paths to evaluate social impacts on tunnel fire evacuation. Andrée et al. (2016) investigated participants' exit choices to assess the usage of elevators in a high-rise building fire evacuation. Cosma et al. (2016) explored the impact of way-finding installations on tunnel evacuation. Together, SGs have been suggested as a promising tool to study human behaviors (Boyle, Hainey, Connolly, Gray, Earp, Ott et al., 2016; Connolly, Boyle, Boyle, MacArthur, & Hainey, 2012). However, while various behavioral measurements exist, little attention has been paid to earthquake emergencies.

SGs have been applied to various platforms, including mobile devices and desktop computers (Connolly et al., 2012). To provide an advanced immersive and engaging experience, SGs can be integrated with Immersive Virtual Reality (IVR), known as IVR SGs. IVR is a technology that can induce a *"targeted behavior in an organism by using artificial sensory stimulation, while the organism has little or no awareness of the interference"* (LaValle, 2017, p. 1). IVR can provide a credible virtual environment where participants can explore and behave as close to reality as possible (LaValle, 2017; Sherman & Craig, 2018). Such real-world reactions are essential for behavioral analysis as well as educational applications as participants are expected to shift their behaviors towards recommended ones after training. Krokos et al. (2018) indicated that participants under an IVR condition had better performance in terms of memory recall as compared to non-IVR conditions. Participants were found to be more focused on tasks when they were fully immersed in the virtual environment provided by IVR. Similarly, Chittaro and Buttussi (2015) argued that IVR was beneficial to knowledge retention because of the highly psychological arousal yielded by the high-degree engagement and life-like experience. The synergies that exist between IVR and SG are apparent, which justify the combination of these approaches.

IVR SGs have become a popular tool for emergency training and research. A recent study by Feng et al. (2018) indicated that IVR SGs had been applied to various emergency situations, including fire evacuation, airplane emergencies, and earthquakes. Smith and Ericson (2009) adopted an IVR SG to increase children's motivation towards learning fire safety skills. The findings of this study revealed that participants improved their fire safety knowledge significantly after the training took place. Burigat and Chittaro (2016) applied an IVR SG to train participants about spatial knowledge of an airplane, in order to undertake an effective evacuation. Participants trained by the IVR SG took less time to evacuate as compared to those trained by safety cards. Li et al. (2017) proposed an IVR SG to train participants in self-protection skills during earthquakes. Participants were asked to detect potential hazards and avoid physical damage during indoor earthquake emergencies. The results suggested that the IVR SG was more effective than other approaches (videos and manuals) in terms of self-protection skills training. Overall, previous studies showed that IVR SGs have the potential to generate positive outcomes for emergency training. However, to date, only a few studies have focused on IVR SGs for earthquake behavioral responses and post-earthquake evacuation preparedness, and paid attention to the education dimension to teach and disseminate best evacuation practice (Feng et al., 2018). Little is known about the effectiveness and applicability of IVR SGs to improve the immediate behavioral responses to earthquakes and post-earthquake evacuation preparedness in buildings.

### 2.2 National and hospital earthquake response procedures in New Zealand

This study took place in New Zealand and Auckland City Hospital (ACH) was chosen as a case study. ACH is New Zealand's largest public hospital and clinical research facility. The reason to use ACH in the IVR SG training system is that hospital evacuation drills are always restricted due to ethical issues and risks from disruptions of operational functions (Johnson, 2006). This gave us the opportunity to run a virtual drill by using IVR SGs. National and hospital earthquake response procedures in New Zealand are reviewed in this section.

New Zealand experiences earthquakes with a frequency of between 150 and 200 perceptible earthquakes a year (McSaveney, 2017). An $M_w$ 6.2 earthquake hit Christchurch in 2011 causing 185 fatalities (New Zealand Police, 2011). As a result, the New Zealand's government has put significant resources into training and educating the general public about the best evacuation practice to respond to earthquake emergencies. Since 2012, and on a yearly basis, the New Zealand's government has promoted a nationwide earthquake drill and tsunami evacuation practice named the New Zealand ShakeOut (New Zealand ShakeOut, 2018). The New Zealand ShakeOut aims to encourage the public to undertake the basic "Drop, Cover and Hold" actions during an earthquake, and to practice a tsunami evacuation if necessary. Drop, Cover and Hold (DCH) has been promoted as the primary action to perform during earthquakes in New Zealand, rather than to pay attention to the Triangle of Life (New Zealand Ministry of Civil Defence & Emergency Management, 2015; Stuart-Black, 2015). DCH encourages people to drop down to maintain balance, take shelter under sturdy furniture (e.g., a table) within a few steps, and hold on to it to maintain protection (Mahdavifar et al., 2009). The Triangle of Life encourages people to curl up next to heavy furniture (e.g., a sofa) in order to obtain a survivable space in case of building collapse (Mahdavifar et al., 2009). DCH is based on the assumption that buildings are not going to have a structural collapse during earthquakes, whereas the Triangle of Life assumes that buildings will collapse and crush furniture inside.

Research from earthquake-prone countries and regions such as the U.S., Taiwan, Japan, Iran, and New Zealand support the recommendation that DCH is the most appropriate action to take in earthquakes (Mahdavifar et al., 2009; Stuart-Black, 2015). After shaking stops, instead of immediately exiting buildings, New Zealand Civil Defence encourages a list of behaviors representing the best evacuation practice for post-earthquake evacuation, such as check surroundings and suitable evacuation pathways, gather important personal items, help others if possible, check for and extinguish small fires if possible, listen to a radio, and evacuate buildings by the stairs (New Zealand Ministry of Civil Defence & Emergency Management, 2015). The general public has various ways to get access to learn the recommended behavioral responses to earthquakes and post-earthquake evacuation. However, they have little chance to practice it. Even in earthquake drills, building occupants often practice DCH only, not the entire set of best practice due to the cost, and the actual hazards and threats are too dangerous to be represented within a real physical setting (Becker, Coomer, Potter, McBride, Lambie, Johnston et al., 2016; Gwynne, Boyce, Kuligowski, Nilsson, P. Robbins, & Lovreglio, 2016; Lovreglio et al., 2017). This supports the notion that IVR SGs have potential as a training tool to promote earthquakes and post-earthquake evacuation preparedness; and therefore, they need to be further investigated.

In addition to the national advice, Auckland District Health Board has issued an Auckland District Health Board Evacuation Plan (Auckland District Health Board, 2009) and an Emergency Preparedness & Response Manual, which include earthquake response procedures for Auckland District Health Board facilities (Auckland District Health Board, 2014). Similar to the recommendations suggested by New Zealand Civil Defence, this Manual encourages hospital staff to DCH during earthquakes and stay in buildings immediately after an earthquake. Also, staff members are recommended to take more responsible actions, such as administer first aid as required, advise visitors to remain until the situation has been assessed for safety, check for and contain hazards such as fire and gas or chemical leaks where practicable, turn off damaged utilities, and unplug unnecessary electrical equipment. Despite having well-defined earthquake response procedures and guidelines, Wabo et al. (2012) found that there was a lack of appropriate mechanisms to effectively implement hospital evacuation plans and properly assess their impacts. In order to address this issue, Johnson (2006) suggested that hospital evacuation plans can be effectively assessed by alternative approaches such as computer simulations. Therefore, IVR SG potentially can be investigated in order to understand how it can provide meaningful insights into the preparedness of hospital occupants during emergency evacuation; and, following that, enhance their preparedness.

## 3. The IVR SG training system

The proposed IVR SG training system allows participants to experience a full indoor earthquakes and post-earthquake evacuation. ACH was selected as a test building. This prototype aims to train ACH occupants to improve their preparedness to cope with earthquake emergencies and acquire knowledge on the best post-earthquake evacuation practice as suggested by New Zealand Civil Defence (New Zealand Ministry of Civil Defence & Emergency Management, 2015) and Auckland District Health Board (Auckland District Health Board, 2014).

### 3.1 Virtual Environment

We chose a portion of the ACH's fifth floor as the training location for our IVR SG training prototype, which covered a public area and an administrative area. This location gave access to different categories of building occupants such as staff, patients and visitors. We used a Building Information Modelling (BIM)-based workflow, which is an idea approach to present dynamic changes such as simulating earthquakes, to develop 3D models for virtual environments (Feng, González, Ma, Al-Adhami, & Mourgues, 2018). The 3D model of this building section was developed using Autodesk Revit (a BIM tool for building modelling, www.autodesk.com). Structural components (e.g., walls, columns, floors) and non-structural components (e.g., furniture, doors, windows, ceiling tiles) were included in this BIM model. The BIM model was then imported into Unity (a game engine with user-friendly interfaces and tools for developing IVR and games, unity3d.com) for IVR and game mechanism development. More details of the workflow can be found in Lovreglio et al. (Lovreglio, González, Feng, Amor, Spearpoint, Thomas et al., 2018). Figure 1 shows the workflow to create a virtual environment for the IVR SG. Figure 2 compares scenes from the virtual model and the real ACH.

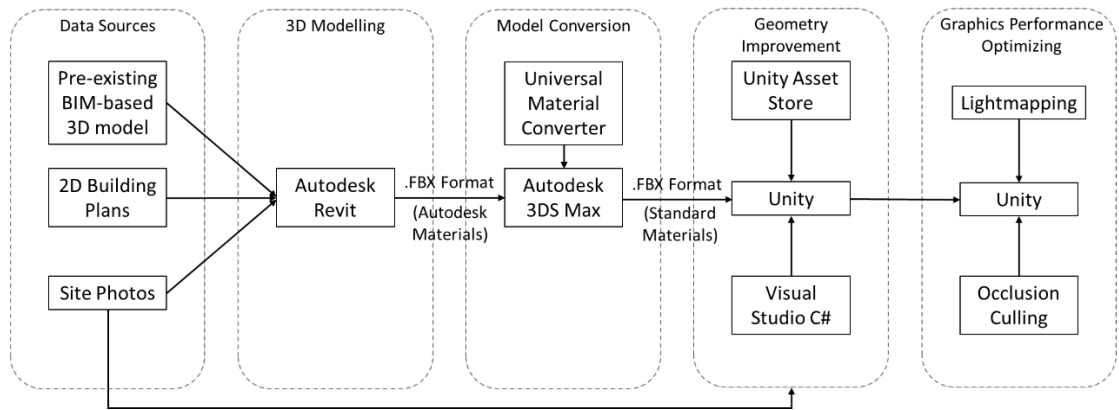

Figure 1 – Workflow to create the virtual environment for the IVR SG (Lovreglio et al., 2018)

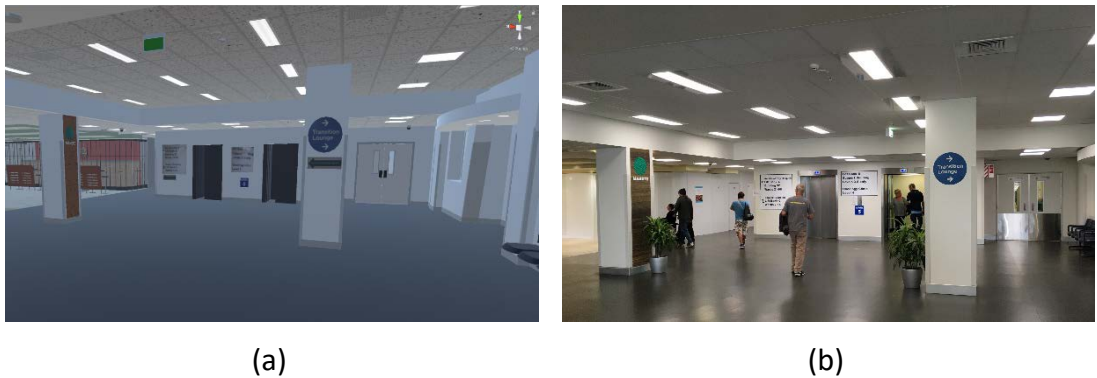

(a)                   (b)

Figure 2 – Comparison between the virtual model (a) and the reality (b) of ACH (Lovreglio et al., 2018)

### 3.2 Earthquake Simulation and Building Damage

We adopted a qualitative strategy to simulate earthquake shaking and damage to the hospital building. This strategy allowed us to mimic damage based on existing datasets of videos and images of building earthquake damage, which excluded the accurate computational simulation of actual structural responses for building elements during and after earthquakes (Lovreglio et al., 2018). The reason to use this strategy is that the purpose of IVR SGs is to provide a training environment with credible and meaningful earthquake and post-earthquake experiences, rather than simulations for structural analysis. In this case, a major failure or collapse of structural components of the hospital building was not considered. We only represented non-structural damage such as falling ceiling tiles, toppling partition walls and furniture, and the breaking of glass panels. Earthquake simulation and building damage were developed using Unity based on the imported BIM model. Figure 3 compares a before-earthquake environment and an after-earthquake environment.

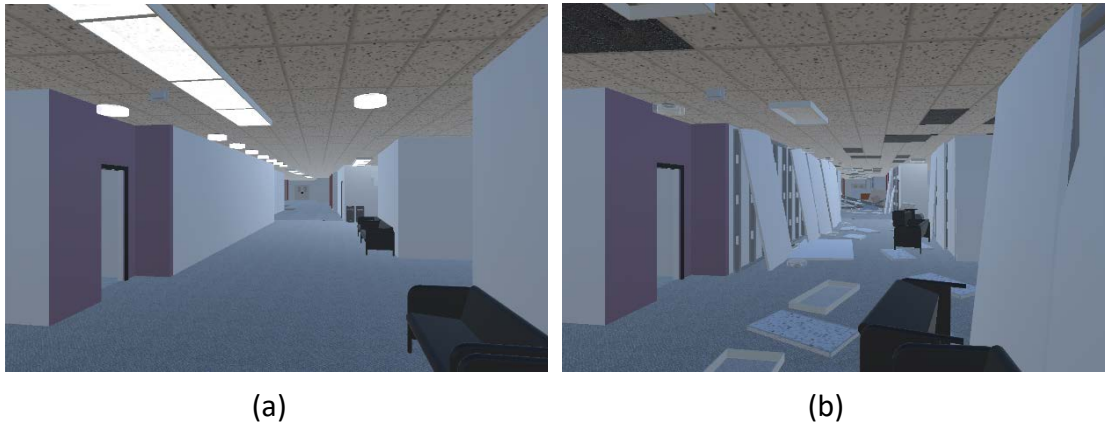

(a)                          (b)

Figure 3 – Comparison between a before-earthquake environment (a) and an after-earthquake environment (b) (Lovreglio et al., 2018)

### 3.3 Non-playable Characters

The IVR SG training system is a single-player game, which means only one participant can play the game at any given time. Non-playable characters (NPCs), which were controlled by predetermined scripts, were therefore introduced to represent other evacuees in our IVR SG. Participants could interact with NPCs in order to perform certain actions (e.g., a female NPC trapped under furniture asks for help as shown in Figure 4).

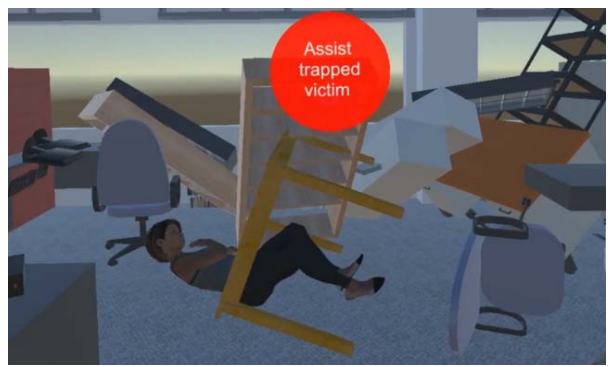

Figure 4 – An example of a female NPC stuck under a table asking for help

### 3.4 Navigation and Interaction

In this IVR SG, we adopted a waypoint system for navigation. Waypoints are sets of coordinates that identify a stopping point or point where navigation routes can be modified (Ragavan, Ponnambalam, & Sumero, 2011, September). Predefined routes were used to connect the waypoints. The navigation was achieved by moving participants' view from a waypoint to another. Participants could turn their bodies to adjust the orientation of their view. This solution limited the participants' movement to prevent them from getting lost or stuck in an open-world IVR environment which occurs if they can move freely, and to reduce motion sickness from the abrupt and non-natural motion of participants (Lovreglio et al., 2018). Whenever participants reached a stopping point, they faced several options that were presented as action panels as shown in Figure 5. These actions were related to the recommended behaviors as listed in Table 1. Participants could make a choice by clicking on one of the panels; and then, they proceed with the stages or scenarios of the IVR SG training system.

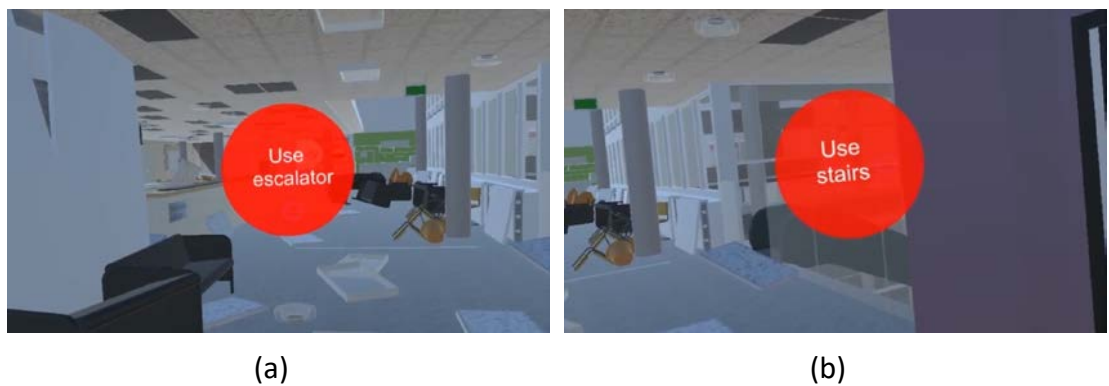

(a)　　　　　　　　　　　　　　　　　　(b)

Figure 5 – An example of two action panels for participants to choose how to exit the building: (a) use an escalator; (b) use a staircase

### 3.5 Training Outcomes and Storyline Narrative

Based on the guidelines provided by New Zealand Civil Defence (New Zealand Ministry of Civil Defence & Emergency Management, 2015) and Auckland District Health Board (Auckland District Health Board, 2014), we identified a list of behaviors as learning outcomes shown in Table 1.

Table 1

Recommended behaviors as learning outcomes

| Phase | Recommended Behaviors |
| --- | --- |
| Indoor Earthquake Phase | Drop, cover and hold |
|  | Pay attention to falling, breaking or dangerous objects around |
| Pre-evacuation and Indoor Evacuation Phase | Stay under cover to see if there are aftershocks |
|  | Collect personal belongings |
|  | Take first aid kit |
|  | Check and help people around |
|  | Search for alternative exits if the closest or usual one is blocked |
|  | Put out a small fire with a fire extinguisher or report it |

|  | to the fire brigade |
|---|---|
|  | Unplug damaged electrical equipment |
|  | Use stairs to exit |
| Outdoor Evacuation Phase | Go to an assembly point (an open space away from buildings and falling objects) |
|  | Do not go back to buildings until it is safe to do so |

As in Feng et al. (2018), we adopted an action-driven narrative method, which means a storyline is driven by a sequence of actions taken by participants. With the waypoint system as described in the previous section, participants were led through different game scenarios in which they needed to choose actions in order to make progress through the storyline. The recommended behaviors were embedded in these game scenarios, which shaped the main storyline of the IVR SG.

The storyline of the IVR SG training system consists of the following points, as shown in Figure 6:

1. Participants start the game outside ACH, and they are asked to reach a meeting room in the hospital by following a staff member.
2. Once participants have reached the meeting room, they are welcomed by a doctor NPC, who invites them to leave their belongings on a table.
3. As they leave their belongings, an earthquake strikes. Participants can choose to take cover under a table, or beside a shelf or a window. If participants do nothing after ten seconds, they will be hit by a falling ceiling tile.
4. When the shaking ends, the doctor in the room leaves to check the situation outside while participants can take several actions available in the scenario. The recommended actions included in this part are:
    i) Stay under cover to see if there are aftershocks;
    ii) Collect their personal belongings;
    iii) Take a first aid kit in the room.
5. Finally, participants have an option to get out of the room to start evacuation. While evacuating participants come across several scenarios in which they can choose whether to take the following actions:
    i) Assist a nurse NPC with an injured victim;
    ii) Help a female NPC trapped under a table;
    iii) Search for an alternative exit if the closer or usual one is blocked;
    iv) Unplug damaged electrical equipment;
    v) Extinguish a small fire or report it to the fire brigade;
    vi) Listen to a radio to collect information;
    vii) Use stairs to exit;
6. Participants reach the exit of the building, and then they can choose a safe assembly point to go.

7. The experience ends in a virtual environment where participants receive a post-game assessment commenting on their behaviors.

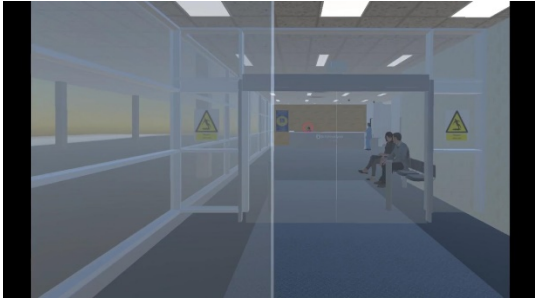

(a)

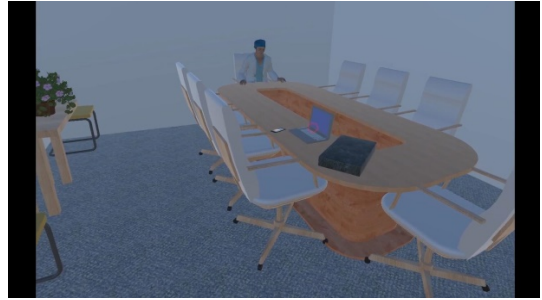

(b)

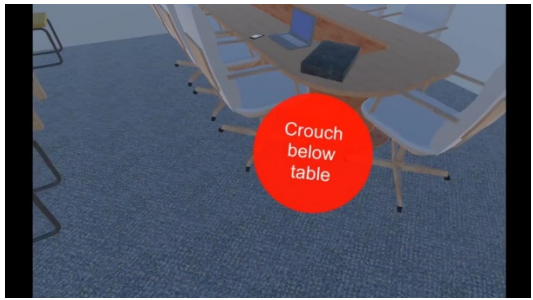

(c)

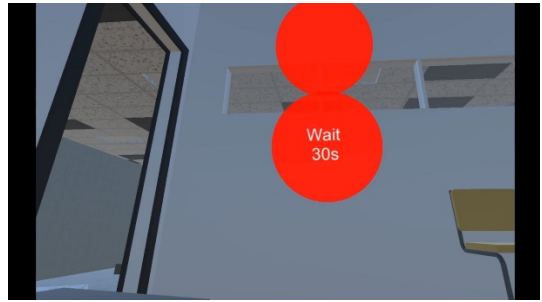

(d)

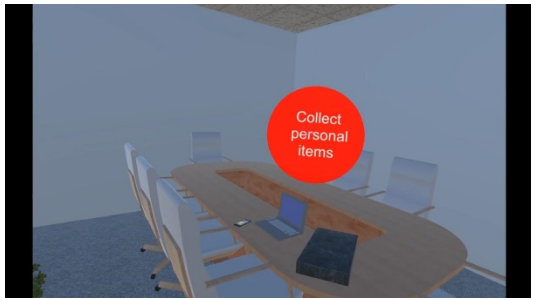

(e)

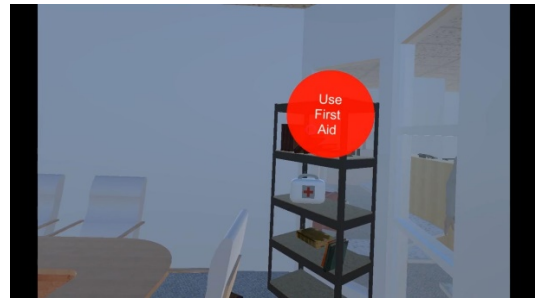

(f)

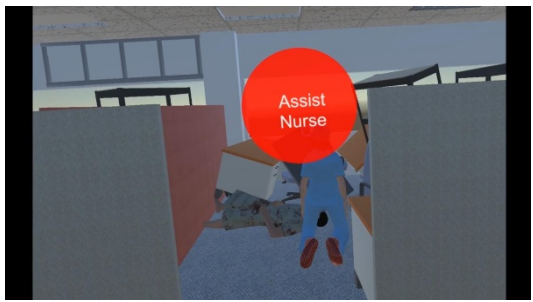

(g)

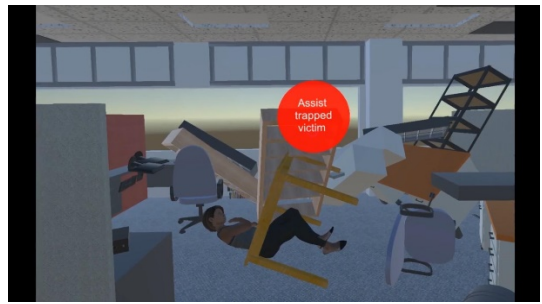

(h)

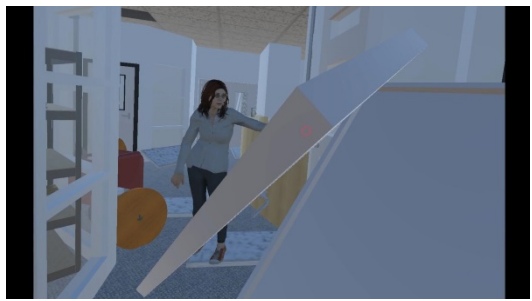

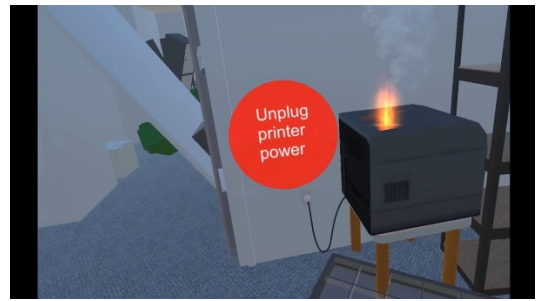

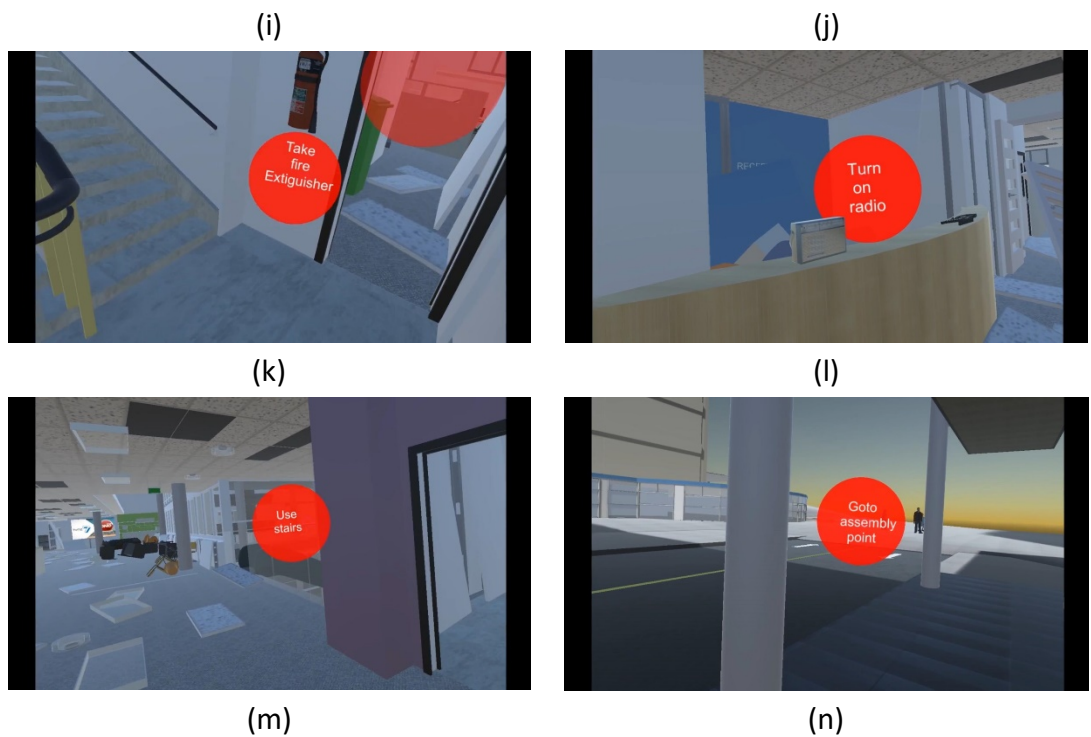

Figure 6 – The storyline of the IVR SG training system: (a) stand outside ACH; (b) welcomed by a doctor NPC in a meeting room; (c) take cover under a table; (d) stay under cover after shaking to see if there are aftershocks; (e) collect personal belongings; (f) take the first aid kit; (g) assist a nurse NPC; (h) help a female NPC; (i) search for alternative exits; (j) unplug a damaged printer; (k) use a fire extinguisher; (l) listen to a radio; (m) use stairs; (n) go to an assembly area in an open space

### 3.6 Teaching Method

Two teaching methods were applied in this training system, namely immediate feedback and post-game assessment (Feng et al., 2018). Regarding immediate feedback, a flashing light was immediately activated after a participant made a choice, indicating whether the decision and further action were recommended or not. If participants chose a recommended action, green lights flashed; whereas for an action that was not recommended, red lights flashed. In this way, participants could immediately receive feedback related to the assessment of their actions. Figure 7 shows an example of flashing green lights indicating participants have chosen a recommended action, and flashing red lights indicating an action that is not recommended.

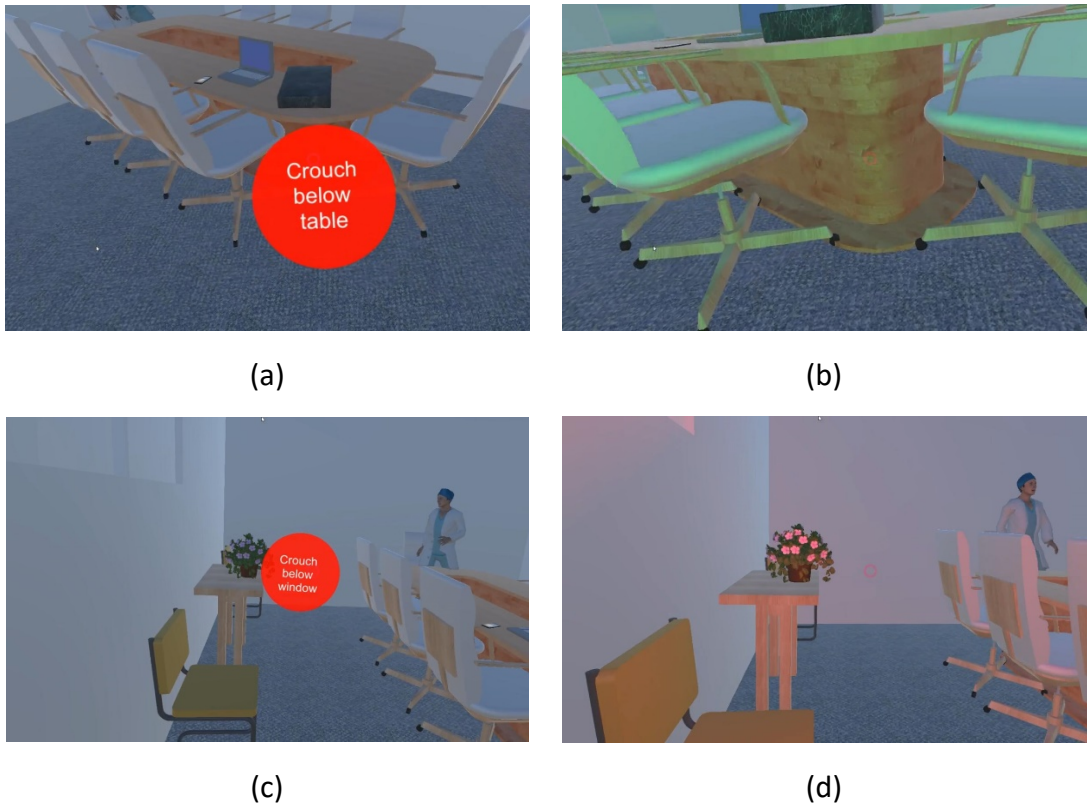

Figure 7 – An example of flashing lights: (a) crouch under a table when an earthquake hits, (b) green lights flash; (c) crouch beside a window, (d) red lights flash

Once the entire evacuation process was over, participants received a detailed post-game assessment, which reported all the actions that had been taken against the full list of recommended behaviors listed in Table 1. Following that, a video and audio playback took participants through all the choices that they made during the training experience, explaining the rationale behind each recommended behavior, as shown in Figure 8. The post-game assessment served as a recap to help participants understand what the recommended behavior was and strengthened their memory to reinforce them.

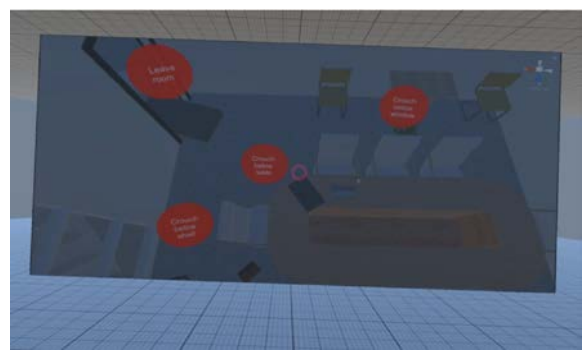

Figure 8 – A screenshot of playback explaining the recommended behaviors

## 4. Research Methods

To evaluate the possible effectiveness and applicability of the IVR SG training system, we carried out a pre-test measure of the outcome of interest prior to administering training, followed by a post-test on the same measure after training occurred. Participants received indoor earthquakes and post-earthquake evacuation training through the IVR SG training system.

### 4.1 Apparatus

The IVR SG training system was implemented as an executable file which was built in Unity. The IVR SG training system was run on a DELL PC workstation equipped with a 2.4 GHz Intel Xeon E5-2640 processor, 64 GB RAM, and two NVidia GTX 1080 graphic cards. The IVR headset was an Oculus Rift (www.oculus.com), which is a head-mounted display (HMD) with 1080x1200 resolution per eye and a 110-degree field of view. The remote controller was an Oculus Remote, which was given to participants to choose action panels by simply pressing one button. The graphic output of the HMD was also displayed on an LED screen, which allowed researchers to observe and record participants' in-game behaviors during the training.

### 4.2 Participants

A total of 93 participants (43 male, 50 female) were recruited to test the IVR SG training system. Participants were contacted by email, leaflets, and posters spread through ACH and The University of Auckland. Of these, 87 participants completed the experiment. The other six had to stop the IVR experience due to motion sickness. The remaining 87 participants consisted of 25 staff members of ACH and 62 visitors. Ages ranged from less than 20 to over 70, with one-third between 20 to 29. The demographic profile is shown in Figure 9.

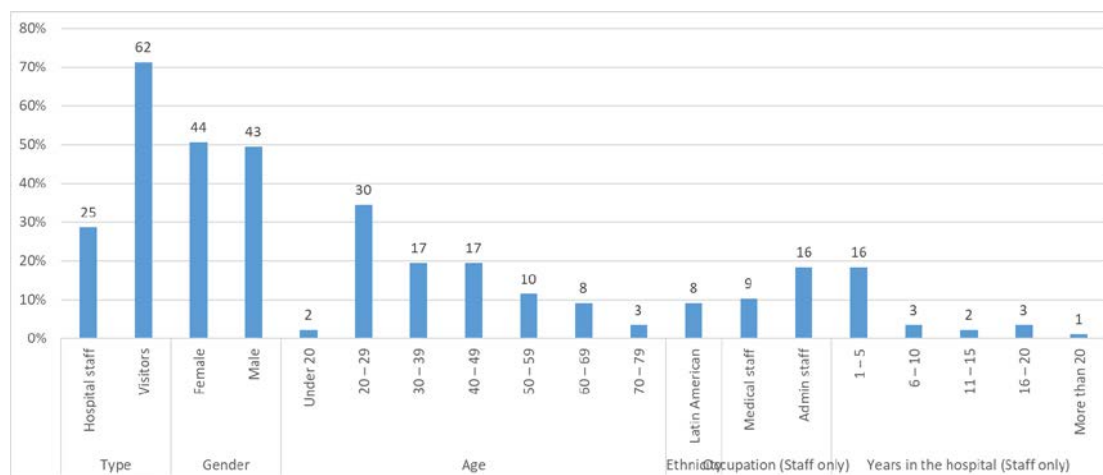

Figure 9 – The demographics of participants

Apart from demographic information, we asked participants about their previous experience involving fire and earthquake drills. Table 2 shows the result of this survey revealing that participants had more experience with fire drills than earthquake drills.

Table 2

Frequencies of practice in fire drills and earthquake drills

| Frequency | Staff | | Visitors | |
| --- | --- | --- | --- | --- |
| | Fire | Earthquake | Fire | Earthquake |
| Never | 7 | 22 | 10 | 45 |
| Once a year | 8 | 1 | 26 | 8 |
| Twice a year | 4 | 0 | 11 | 0 |
| More than twice a year | 3 | 1 | 9 | 1 |
| Unsure | 3 | 1 | 6 | 5 |
| Other | 0 | 0 | 0 | 3 (Only once) |

We also asked participants to state how often they play video games, if at all. Table 3 shows the result of this survey.

Table 3

Frequencies of playing video games

| Frequency | Staff | Visitors |
| --- | --- | --- |
| Never | 11 | 19 |
| Less than once a year | 3 | 17 |
| At least once a year | 5 | 9 |
| At least once a month | 2 | 4 |
| At least once a week | 1 | 5 |
| Several days a week | 2 | 3 |
| Every day | 1 | 5 |

Finally, we asked participants to state whether they had experienced IVR (e.g., games, videos, tours, or demos) before. Table 4 shows the result of this survey.

Table 4

Experience with IVR

| Experience | Staff | Visitors |
| --- | --- | --- |
| No | 13 | 31 |
| Yes | 11 | 23 |

| Unsure | 1 | 8 |
|---|---|---|

### 4.3 Measures

Individuals' preparedness for earthquakes and post-earthquake evacuation can be influenced by various factors such as hazard perception, attitude, and experience (Tekeli-Yeşil, Dedeoğlu, Tanner, Braun-Fahrlaender, & Obrist, 2010). This study deployed two instruments to assess participants' preparedness, namely knowledge about proper behavioral responses and self-efficacy in dealing with such emergencies. Besides, this study measured self-report training efficacy and engagement in order to give insights about the applicability of the proposed IVR SG training system.

#### 4.3.1 Knowledge acquisition

In order to measure participants' knowledge associated with the immediate behavioral responses to earthquakes and post-earthquake evacuation, we asked five questions. These five questions were focused on three aspects, namely knowledge of behavioral responses: inside a building during an earthquake, inside a building after an earthquake, and outside a building after an earthquake. The questions were open-ended questions in order to avoid the suggestion of possible answers that could bias responses from participants. Participants answered the same five questions orally before and immediately after the training, and their answers were audio recorded. Each recorded audio was transcribed and coded by three researchers. For each assessed aspect, three transcriptions from three researchers were cross-checked and merged into one final transcription. According to the final transcriptions, scores were given based on a knowledge scale. The knowledge scale was developed based on the recommended behaviors that were identified as learning outcomes in Table 1. Possible scores ranged from 1 to 4, where 1 stood for no knowledge and 4 stood for strong knowledge. As a result, every participant received three scores for pre-training and three scores for post-training. Table 5 shows the assessed aspects, open-ended questions, and knowledge scale.

Table 5

Assessed knowledge aspects, open-ended questions for pre- and post-training, and knowledge scale

| Knowledge aspects | Asked questions | Knowledge scale | | | |
|---|---|---|---|---|---|
| | | Strong knowledge (4 points) | Adequate knowledge | Weak knowledge | No knowledge (1 point) |

| | | | (3 or 3.5 points) | (2 or 2.5 points) | |
|---|---|---|---|---|---|
| Behavioral responses inside a building during an earthquake | Q1: What would you do during an earthquake? | 4 points for knowing to (i) drop, cover and hold under a table, (ii) pay attention to falling or unsteady objects and glass. | 3 points for knowing to drop, cover and hold under a table | 2 points for knowing to take cover or find a shelter. | 1 point for knowing nothing. |
| | Q2: What should you pay attention to during an earthquake? | | | | |
| Behavioral responses inside a building after an earthquake | Q3: What would you do after an earthquake? | 4 points for knowing over eight items out of eleven items as stated below (i) stay undercover to see if there are aftershocks, (ii) collect personal items, (iii) pay attention to first aid kit, (iv) pay attention to people around and offer help, (v) search for available exits if common ones are blocked, (vi) pay attention to fire, (vii) pay attention to fire extinguishers, put out fire if practicable or call fire departments, (viii) pay attention to electric leakage, (ix) unplug equipment if practicable, (x) listen to a radio to get more information and instructions, (xi) use stairs to exit buildings. | 3 points for knowing five or six items as specified in the strong knowledge column; 3.5 points for knowing seven or eight items. | 2 points for knowing one or two items as specified in the strong knowledge column; 2.5 points for knowing three or four items. | 1 point for knowing nothing. |
| | Q4: What should you pay attention to after an earthquake? | | | | |
| Behavioral responses outside a building after an earthquake | Q5: What is the correct behavior when you are outside a building after an earthquake? | 4 points for knowing to (i) stay at an open space which is away from buildings and falling objects, (ii) don't go back to buildings until it's safe to do so. | 3 points for knowing to stay at an open space which is away from the buildings and falling objects. | 2 points for knowing to go to an assembly point only. | 1 point for knowing nothing. |

### 4.3.2 Self-efficacy

Self-efficacy is a person's belief in his or her ability to successfully accomplish difficult tasks (Bandura, 1982). Self-efficacy can largely influence a person's behavior and performance outcomes (Bandura, 1977; Stajkovic & Luthans, 1998). In order to measure participants' levels of self-efficacy in dealing with earthquake emergencies, we administered a questionnaire before and immediately after the training. The questionnaire was designed based on the General Self-Efficacy Scale (Schwarzer & Jerusalem, 2010). We asked participants to rate their levels of agreement on a 7-point Likert scale (-3 = strongly disagree, +3 = strongly agree) about six statements:

1. "I am confident that I am able to effectively deal with an earthquake emergency";
2. "Thanks to my resources, I know how to manage in an earthquake emergency";
3. "I would be able to deal with an earthquake emergency even if the building is severely damaged";
4. "I would be able to deal with an earthquake emergency even if I find flame and fire along the way";
5. "I would be able to deal with an earthquake emergency even if the exit is blocked";
6. "I would be able to deal with an earthquake emergency even if I find objects that may harm me along the way."

### 4.3.3 Self-reported training efficacy

To measure the levels of training efficacy perceived by participants, we administered a questionnaire immediately after the training. We asked participants to rate their levels of agreement on a 7-point Likert scale (-3 = strongly disagree, +3 = strongly agree) about the following statement: "I could easily learn the recommendations provided in the virtual game."

### 4.3.4 Self-reported engagement

To measure the levels of engagement experienced by participants, we administered a questionnaire immediately after the training. We asked participants to rate their levels of agreement on a 7-point Likert scale (-3 = strongly disagree, +3 = strongly agree) about this statement: "The game was engaging/fun."

### 4.4 Procedure

The experiment was carried out in a meeting room at ACH. Participants received participation information sheets, which informed them that this experiment aimed to test an IVR SG training system that was designed for an earthquake emergency. Participants then received consent forms, which requested consent for participation and collecting research data including questionnaires, audio recordings, and in-game actions recordings by the researchers. Participants were informed that they could stop and quit the experiment at any time without giving any reason.

After signing the consent forms, participants were asked to fill in questionnaires including demographic information, frequency of practice in fire drills and earthquake drills, frequency of playing video games, experience with IVR, and self-efficacy in dealing with an earthquake emergency. Following that, participants were asked to orally answer a five-question knowledge test for pre-training knowledge assessment.

After pre-training knowledge assessment, participants were given introductions about controls as well as health and safety for using IVR. Then, participants were invited to wear an HMD and adjust it until they had a clear view and felt comfortable with it. Participants were asked to sit in a swivel chair for the entire IVR SG training session, which made it easy to turn their bodies with minimum falling risk. Before participants were exposed to the IVR SG training session, they were led through a tutorial session (an empty room with waypoints and action panels) that helped participants understand the navigation and interaction of the IVR SG training system and get familiar with IVR environments and controls. Then, the training session was started. The entire training session was around 20 minutes.

After the IVR SG training session was completed, participants orally answered the same five-question knowledge test again for post-training knowledge assessment. Then, participants filled in questionnaires about self-efficacy, training efficacy, and engagement.

## 5. Results

### 5.1 Knowledge acquisition

A Shapiro-Wilk test indicated that the knowledge scores for pre- and post-training were non-normally distributed for both staff and visitors. Given that, a non-parametric test was suitable for pairwise comparisons. A Wilcoxon Signed-rank test indicated that for both staff and visitors, their knowledge levels of behavioral responses regarding three situations had increased significantly after the training, as shown in Table 6. Boxplots were also used to show the comparisons of knowledge scores, as shown in

Figure 10.

Table 6

Wilcoxon Signed-rank test results for knowledge levels comparisons

| Knowledge Aspects | Staff | | Visitors | |
|---|---|---|---|---|
| | Pre-training | Post-training | Pre-training | Post-training |
| Behavioral responses inside a building during an earthquake | M = 2.44<br>SD = 1.16<br>Z = -2.452, p = 0.014 | M = 3.08<br>SD = 1.00 | M = 2.47<br>SD = 1.26<br>Z = -3.092, p = 0.002 | M = 3.00<br>SD = 0.99 |
| Behavioral responses inside a building after an earthquake | M = 1.90<br>SD = 0.43<br>Z = -3.745, p = 0.000 | M = 2.68<br>SD = 0.64 | M = 1.87<br>SD = 0.46<br>Z = -6.135, p = 0.000 | M = 2.63<br>SD = 0.57 |
| Behavioral responses inside a building after an earthquake | M = 2.20<br>SD = 1.08<br>Z = -2.168, p = 0.030 | M = 2.68<br>SD = 0.63 | M = 2.37<br>SD = 0.94<br>Z = -2.995, p = 0.003 | M = 2.74<br>SD = 0.68 |

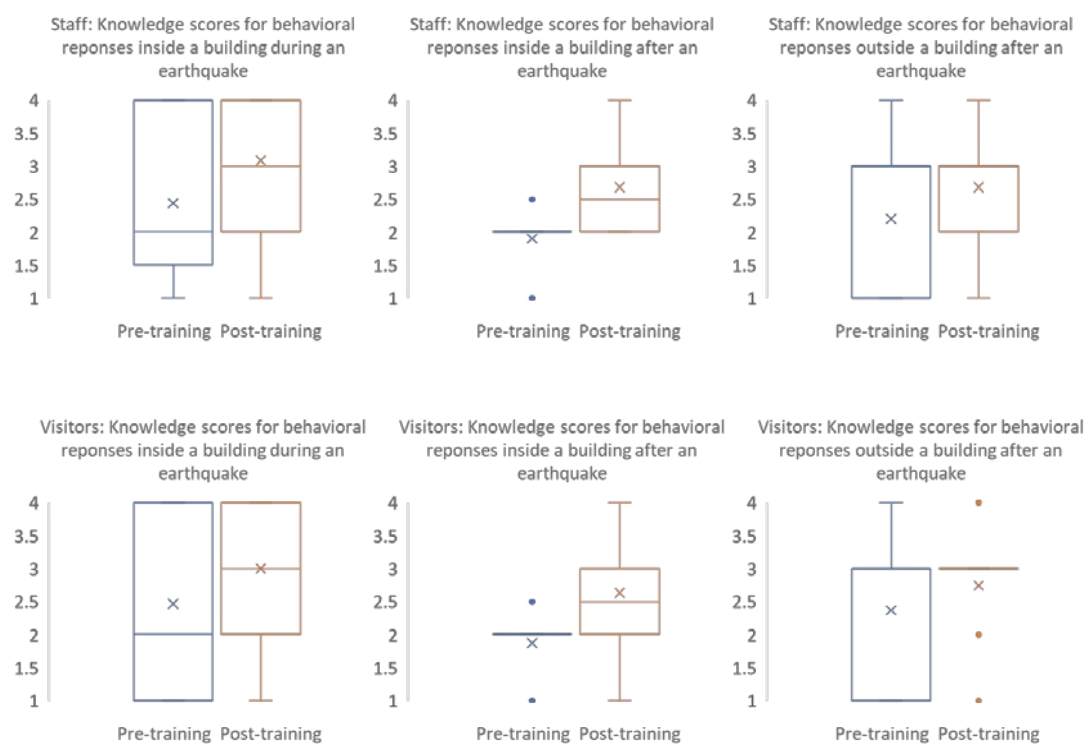

Figure 10 – Comparisons of participants' knowledge scores using boxplots

## 5.2 Self-efficacy

The regression method of factor analysis was used to estimate a factor score for each participant based on the answers from six statements. A Shapiro-Wilk test indicated that the staff's estimated factor scores for pre-training were normally distributed, while their scores for post-training and visitors' scores for pre- and post-training were non-normally distributed. Given that, a non-parametric test was suitable for pairwise comparisons. A Wilcoxon Signed-rank test indicated that for both staff and visitors, their self-efficacy had increased significantly after the training, as shown in Table 7. Boxplots were also used to show the comparisons of estimated factor scores for self-efficacy, as shown in Figure 11.

Table 7

Wilcoxon Signed-rank test results for self-efficacy levels comparisons

| Staff | | Visitors | |
| --- | --- | --- | --- |
| Pre-training | Post-training | Pre-training | Post-training |
| M = -0.61 | M = 0.61 | M = -0.69 | M = 0.69 |
| SD = 0.85 | SD = 0.70 | SD = 0.94 | SD = 0.59 |
| Z = -4.286, p = 0.000 | | Z = -6.587, p = 0.000 | |

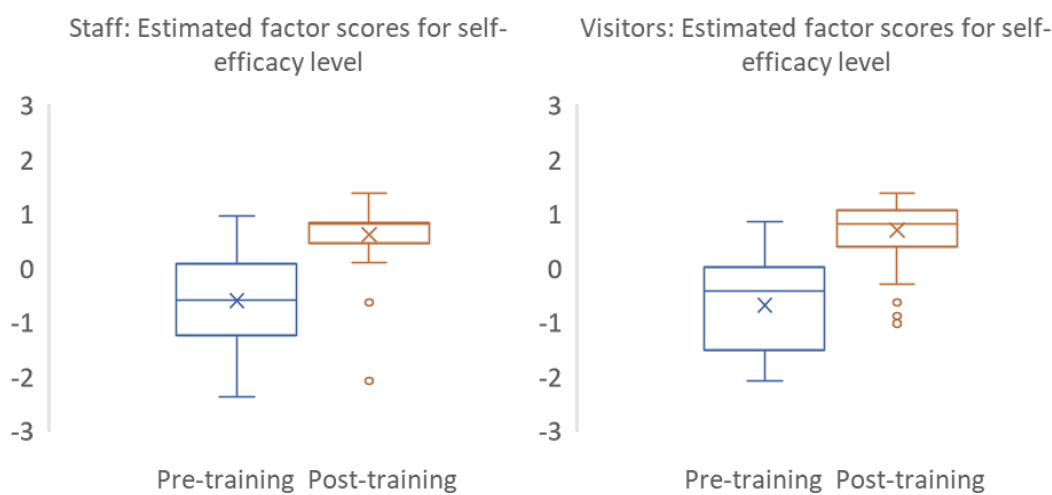

Figure 11 - Comparisons of self-efficacy levels using boxplots

### 5.3 Self-reported training efficacy

The result of training efficacy rating was reported as boxplots (Staff: M = 2.52, SD = 0.59; Visitors: M = 2.65, SD = 0.66), as shown in Figure 12. The result indicated that participants agreed that the IVR SG facilitated their learning immediate behavioral responses to earthquakes and post-earthquake evacuation.

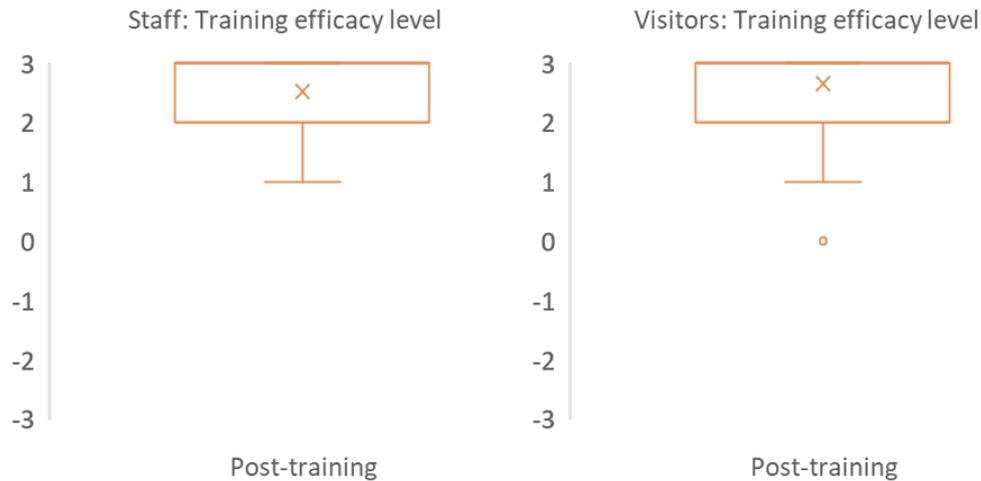

Figure 12 – Perceived training efficacy by participants

### 5.4 Self-reported engagement

The result of engagement rating was reported as boxplots (Staff: M = 2.28, SD = 0.79; Visitors: M = 1.79, SD = 1.26), as shown in Figure 13. The result showed that participants perceived that the IVR SG was engaging.

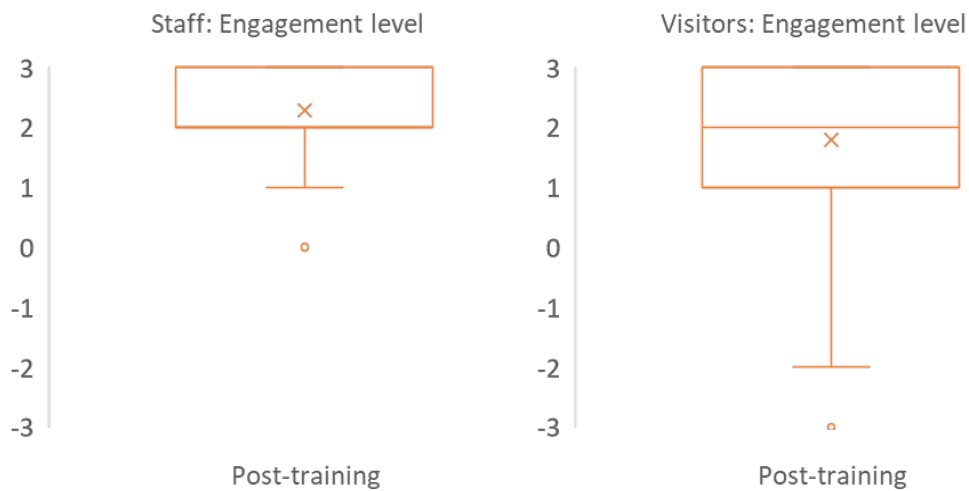

Figure 13 – Perceived usability by participants

### 6. Discussion

This study provides insights about the effectiveness and applicability of IVR SGs used to improve individuals' preparedness for indoor earthquakes and post-earthquake evacuation. The results of the knowledge test in Figure 11 show that participants had a significant increase in knowledge about behavioral responses during earthquakes and post-earthquake evacuation immediately after the training. The results of self-efficacy measurement in Figure 12 indicate that participants had a significant improvement in their confidence about their ability to deal with earthquakes and post-earthquake evacuation after the training. Taken together, these results suggest that the proposed IVR SG training system is effective in increasing an individual's preparedness for earthquake emergencies. The results of training efficacy and engagement measurements in Figure 13 and 14 reveal that the IVR SG training system is applicable, engaging and it facilitates teaching behaviors.

IVR SGs have been widely adopted for emergency training (Feng et al., 2018). Emergency training targeting various emergency types have different levels of training workloads for participants. Studies using IVR SGs have allowed for the training of single behavioral responses such as self-protection skills during earthquakes (Li et al., 2017) and multiple behavioral responses such as the best practice for aviation emergencies (Chittaro & Buttussi, 2015). Our study introduces 13 behavioral responses based on the recommendations from New Zealand Civil Defence and Auckland District Health Board, which were identified as learning outcomes for the IVR SG training system, as shown in Table 1. These behavioral responses made up earthquakes and post-earthquake evacuation procedures in three phases, namely: indoor earthquake phase, pre-evacuation and indoor evacuation phase, and outdoor evacuation phase. Participants increased their knowledge significantly after the training in all three phases. This finding is consistent with that of Chittaro and Buttussi (2015) who pointed out that their IVR SG was effective in increasing safety knowledge including ten behavioral responses on aviation emergencies. Taken together, these findings show that IVR SGs have the potential to effectively deliver complex training and learning content in terms of the best practice for emergency responses.

While IVR SGs are suggested to be effective to train participants about multiple behavioral responses with certain workloads, different performances in our training outcomes are apparent. As shown in Table 8, knowledge of behavioral responses inside a building during an earthquake achieved the highest scores both in pre- and post-training as compared to the other two assessed knowledge aspects. One possible factor that can contribute to this is that we assessed eleven items in knowledge on behavioral responses inside a building after an earthquake. Participants might give different degrees of detail when answering knowledge test questions since open-ended questions heavily rely on the effort from respondents; therefore, this could lead to an inaccurate assessment (Vinten, 1995). Another possible explanation for this could be that participants had a lack of awareness of what to do immediately after earthquakes. Which is evidenced by other recent earthquakes, where behaviors like freezing in place (as opposed to drop, cover, hold) is still the most common behavior (Lindell, Prater, Wu, Huang, Johnston, Becker et al., 2016). Participants need more effort to gain and retain knowledge about other behavioral responses. In order to address this issue, we believe that multiple practices with different training environments can be applied to participants. As Steven (1982) argued, memory recall can be enhanced by using multiple environmental contexts during learning processes. This is aligned with Chittaro and Sioni's (2015) suggestions that repetitive rehearsals can be introduced to improve learning outcomes of IVR SGs.

Table 8

Pre- and post-training knowledge scores for all participants

| Knowledge Aspects | Pre-training | Post-training |
| --- | --- | --- |
| Behavioral responses inside a building during an earthquake | M = 2.46<br>SD = 1.23 | M = 3.02<br>SD = 0.99 |
| Behavioral responses inside a building after an earthquake | M = 1.88<br>SD = 0.45 | M = 2.64<br>SD = 0.59 |
| Behavioral responses outside a building after an earthquake | M = 2.32<br>SD = 0.98 | M = 2.72<br>SD = 0.66 |

In Section 5.1, we split knowledge acquisition analysis between staff and visitors. However, there were no considerable differences identified within these two groups. One possible reason is that the selected behaviors as learning outcomes from the ACH evacuation plan are also included by the national earthquake response procedures. We did not explicitly include particular behaviors in our IVR SG training system for hospital staff. The training content was generally applicable to both staff and visitors. This might lead to the result that staff and visitors had similar performance regarding knowledge acquisition.

Apart from knowledge acquisition, our study also measured self-efficacy as another factor to influence earthquakes and post-earthquake evacuation preparedness. Self-efficacy is a personal belief and an important predictor of attitude and behavior change (Bandura, 1977). Our study shows that participants increased their self-efficacy significantly after the training. This finding is in agreement with Chittaro and Sioni's (2015) findings, which showed that SGs were effective in increasing individuals' self-efficacy in terror attack emergency preparedness. Chittaro and Sioni (2015) argued that SGs provided effective actions for participants to choose when they were threatened by risks. In this way, it was beneficial to increase self-efficacy since participants were motivated to take actions to protect themselves (Chittaro & Sioni, 2015). IVR SGs have the potential to provide engaging environments for participants to go through life-like hazards and be trained to respond to them effectively. As a result, participants are motivated to be more confident facing actual earthquake emergencies and to perform better when dealing with such emergency situations (Stajkovic &

Luthans, 1998).

In terms of applicability, participants reported a high level of perceived training efficacy and engagement. This finding indicates that participants felt engaged and the knowledge learning process was easy with the IVR SG training system, which is in accordance with previous studies (Chittaro & Sioni, 2015; Smith & Ericson, 2009). Participants from Chittaro and Sioni's (2015) were students from universities (Mean age = 23.68), while participants from Smith and Ericson's (Smith & Ericson, 2009) were children aged from seven to eleven. In our study, 87 final participants were mainly adults, with eleven of them being between 60 to 79 years old. And over half of our participants had never experienced IVR before. Surprisingly, as an innovative digital technology, IVR SG seems to be well accepted and easy to use as an emergency training tool for various age groups. However, we must note that six out of 93 participants had to quit our experiment due to motion sickness caused by the IVR experience. A deeper understanding of how to improve the design of IVR SGs to promote user-friendly, no motion sickness training environments for different age groups would be beneficial to influence a larger audience. This is especially important for children at school age as they are receptive to new knowledge and play an important role for a community to build up disaster prevention culture, as highlighted in previous studies (Bernhardsdottir, Musacchio, Ferreira, & Falsaperla, 2016; Izadkhah & Hosseini, 2005; Shaw, Kobayashi, & Kobayashi, 2004). Further studies can pay more attention to this point.

## 7. Conclusions

There has been little research on IVR SGs for earthquakes, especially focused on the full behavioral responses for earthquakes and post-earthquake evacuation. In order to fill this gap, we investigated the effectiveness and applicability of an IVR SG as a training tool to enhance the immediate behavioral responses to earthquakes and post-earthquake evacuation preparedness. We have shown that the proposed IVR SG training system was effective to enhance an individual's preparedness for earthquakes and post-earthquake evacuation. Participants' knowledge of behavioral responses and self-efficacy increased significantly after the training. Additionally, we have shown that the IVR SG was engaging and easy to use for learning immediate behavioral responses to earthquakes and post-earthquake evacuation. This demonstrates that although novel, IVR SG has the potential to be applied as a robust tool for emergency response training.

A topic that was not addressed is whether an IVR SG helps to achieve better learning outcomes as compared to other traditional training approaches such as seminars, posters, and videos. Knowledge retention assessment was also not carried out in this study to see how well participants retained the newly grasped knowledge through IVR SGs over time.

Potential directions for future studies are also identified. Most of the participants in our study were young adults. How IVR SGs fit for children and seniors in terms of earthquakes and post-earthquake evacuation training remains unclear. Greater effort should be made to mitigate motion sickness caused by an IVR experience, given that around 6% of participants had to stop the experiment after suffering from it.

## Acknowledgments

This research has been funded by the MBIE-Natural Hazards Research Platform (New Zealand), Grant Number: C05X0907. The authors thank Andrew George, Bunpor Taing, Hakshay Kumar, and Matthew Richards, undergraduate students in Civil Engineering at The University of Auckland, for the development of the ACH BIM model; Bruce Rooke for the software development; and Gujun Pu, Mohammed Adel Abdelmegid, Ashkan Mohajeri Naraghi, Rehan Masood, and Saleh Alazmi, Ph.D. students in Civil Engineering at The University of Auckland, for data collection and analysis.